%% file: main.tex
\documentclass{article}

\usepackage{amsmath,graphicx}
\usepackage{multirow}
\usepackage{caption}
\usepackage{subcaption}
\usepackage{tipa}
\usepackage{booktabs}
\usepackage[preprint]{spconf}

\title{The Use of Voice Source Features for Sung Speech Recognition}
\name{Gerardo Roa Dabike, Jon Barker}

\address{
  Department of Computer Science, University of Sheffield, UK}

\begin{document}
\ninept
\maketitle
\begin{table}[t]

\copyright\ 2021 IEEE. Personal use of this material is permitted. Permission from IEEE must be obtained for all other uses, in any current or future media, including reprinting/republishing this material for advertising or promotional purposes, creating new collective works, for resale or redistribution to servers or lists, or reuse of any copyrighted component of this work in other works.
\end{table}

\copyrightnotice{\copyright\ IEEE 2021}
                 \toappear{To appear in {\it Proc.\ ICASSP 2021,
                   June 06-11, 2021, Toronto, Ontario, Canada}}

\input{0-abstract}

\input{1-introduction}



\input{2-background}

\input{3-pitchtracker}

\input{4-baseline_system}

\input{5-experiments}

\input{6-conclusion}

\bibliographystyle{IEEEtran}

\bibliography{IEEEabrv, mybib}

\end{document}

%% file: 0-abstract.tex
\begin{abstract}

In this paper, we ask whether vocal source features (pitch, shimmer, jitter, etc) can improve the performance of automatic \textit{sung} speech recognition, arguing that conclusions previously drawn from spoken speech studies may not be valid in the sung speech domain. We first use a parallel singing/speaking corpus (NUS-48E) to illustrate differences in sung vs spoken voicing characteristics including pitch range, syllables duration, vibrato, jitter and shimmer. We then use this analysis to inform speech recognition experiments on the sung speech DSing corpus, using a state of the art acoustic model and augmenting conventional features with various voice source parameters. Experiments are run with three standard (increasingly large) training sets, DSing1 (15.1 hours), DSing3 (44.7 hours) and DSing30 (149.1 hours). Pitch combined with degree of voicing produces a significant decrease in WER from 38.1\% to 36.7\% when training with DSing1 however smaller decreases in WER observed when training with the larger more varied DSing3 and DSing30 sets were not seen to be statistically significant. Voicing \textit{quality} characteristics did not improve recognition performance although analysis suggests that they do contribute to an improved discrimination between voiced/unvoiced phoneme pairs.

\end{abstract}

%% file: 1-introduction.tex
\section{Introduction}


Automatic sung speech recognition is attracting increased research attention, driven in part by the release of large sung speech datasets, e.g., in 2020, the Music Information Retrieval Exchange (MIREX) ran its first Lyrics Transcription task\footnote{https://www.music-ir.org/mirex/wiki/2020:Lyrics\_Transcription}. Lyric transcription is an interesting problem in its own right, but equally importantly, adapting speech technology to sung speech may also provide insight into how to extend to other types of atypical speech (e.g., dysarthric speech). Further, poorly-intelligible sung speech acts as a useful stress-test for state-of-the-art (SOTA) acoustic modelling (AM) techniques.

Existing automatic sung speech recognition systems are typically based on successful approaches for spoken speech \cite{Mesaros2010, Kruspe2016, Tsai2018}. In particular, they use the same acoustic features.
This has been motivated by the idea that spoken and sung speech share the same production system, and that semantic information is conveyed in the same way in both speech styles.
However, there are several differences between sung and spoken speech that make the former more difficult to recognise, such as, the pitch ranges, the syllable duration, the existence of vibrato in singing and the fundamental fact that in singing, intelligibility is often less important than artistic interpretation.
Since most of these differences centre around the use of the voice source, informing recognition systems with voice source information may help to improve their performances.

Traditional SOTA ASR systems designed for spoken speech, typically use Mel frequency cepstral coefficients (MFCC) or filterbank acoustic features to capture the vocal filter characteristics, and i-Vector speaker-specific representations \cite{Dehak2011} for speaker \cite{Panayotov2015, Kyu2018} and environment adaptation \cite{Rouvier2014}. These systems produce excellent results.
However, unaccompanied sung speech ASR scenarios systems that utilised a combination of MFCCs and i-Vectors have not been achieving the same level of performance \cite{Roa2019, Tsai2018}.
This reduced performance may be partly explained by the relative lack of training data availability, the utilisation of inappropriate language models, and the (often) lower intelligibility of the signal.
Additionally, sung speech possesses a higher pitch variability than spoken speech, e.g., females average 342 Hz when singing and 237 Hz speaking (Section \ref{sec:background}).
This variability means that one singer can effectively have different `voices' (guided by the song structure), which may make it difficult to characterise the speakers adequately.


The inclusion of pitch information has been improving the performance of ASR systems in different spoken speech scenarios.
In tonal languages systems, such as Mandarin \cite{Huang2017}, it has been widely used since the tone has a direct relation with the word meaning. 
In non-tonal languages systems, like English \cite{Doss2003, Cloarec2006, plonkowski2016use}, the pitch-strength information helps to improve the discrimination between voiced and unvoiced sounds \cite{Doss2003, Cloarec2006}.
Additionally, pitch information has been used for vocal tract normalisation (VTN), by exploiting the relationship between the pitch values and the vocal tract length, e.g., male speakers have a larger vocal tract than females and produce a lower pitch \cite{plonkowski2016use}.
For children's ASR systems, pitch informed VTN has also been employed to train on adult speech data and testing in children data, by pitch-adaptive front-end \cite{Shahnawazuddin2016, Shahnawazuddin2017}.

However, the role of the voice source in singing is fundamentally different from its role in speaking, meaning it has very different characteristics.
First, for sung speech, the pitch range is larger, and the average pitch is higher than in spoken speech \cite{Merrill2017}. Second, the spoken speech pitch varies freely rising and falling within one syllable, with changes up to 12 semitones \cite{Patel2008}.
In contrast, the pitch in sung speech is expected to remain steady during a note, having controlled and discrete variations, with rather infrequent changes greater than two semitones \cite{vos1989}. Third, in singing, in order to carry the melody, speech is more heavily voiced, and this may interfere with the role of voicing as a phonetic cues. This may lead to difficulties in reliably discriminating between voiced/unvoiced phoneme pairs. Fourth, the duration of the sung vowels is often larger than in spoken speech, as needed to achieve the rhythm dictated by the musical composition and to convey the pitch assigned to the syllable. 
Finally, for artistic expression, singers employ a frequency modulation called vibrato, a musical effect corresponding to small and periodic variations of the pitch between 5.5 and 7.5 Hertz (Hz), with characteristics that are specific to the singer \cite{Sundberg1987Tsotsv}.


Considering these differences, the voice source features may be less useful as a phonetic cue for sung speech than for spoken speech.
Alternatively, with sufficient data, voiced source features, and in particular, pitch, may be, on balance, beneficial for sung speech ASR systems, not only by improving the voiced and unvoiced sounds discrimination and sung vowel classification \cite{Kawai2017}, but also by helping the system to normalise for different speakers or for systematic changes in phoneme quality apparent when a singer is singing in a different part of his or her vocal range.
Additionally, using characterisation of the voice source \textit{quality}, which are strongly associated with speaker identity, may also help the system by providing conditioning variables that improve speaker normalisation.


This paper aims to critically evaluate the usefulness of various types of voice source information, for improving the performance of sung speech ASR systems. This is the first study using a system tested and trained on large amounts of sung speech data. (Pitch information was previously evaluated in \cite{Kawai2017} but in a system trained using spoken data and evaluated only with male singers). Section \ref{sec:background} presents a data-driven analysis of some of the differences between sung and spoken speech.
Section \ref{sec:musicalfeatures} details the voice source-based features chosen for use in the ASR experiments, informed by this analysis.
Section \ref{sec:baseline}, describes the speech recognition dataset and the baseline system.
The experimental results and analysis are presented in Section \ref{sec:experiments} and conclusions in Section \ref{sec:conclusion}.

%% file: 2-background.tex
\section{Sung speech analysis}
\label{sec:background}



Using the sung and spoken lyrics corpus NUS-48E \cite{Duan2013}, we examined some of the sung speech characteristics that makes it more challenging than spoken speech.
The NUS-48E corpus is composed of 48 sung and spoken (read) English songs, performed by 12 singers with different levels of musical training, where each singer speaks and sings four different songs from a selection of 20 unique songs.
The data is humanly annotated at the phone-level, and has about 2800 annotated phones per subject, for each speech style.

\begin{figure}[!t]
    \centering
    \begin{subfigure}{.5\linewidth}
        \centering
        \includegraphics[width=\linewidth]{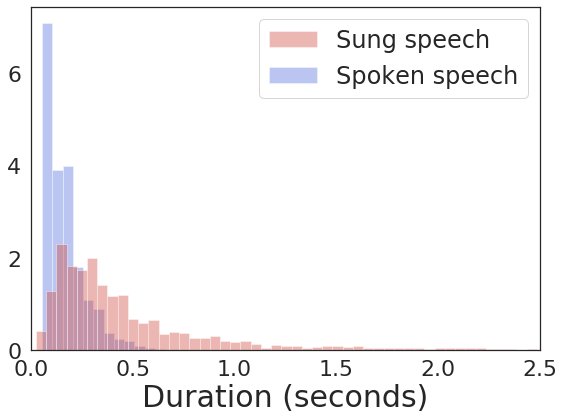}
        \caption{Vowels duration.}
        \label{fig:distDuration}
    \end{subfigure}%
    \begin{subfigure}{.5\linewidth}
        \centering
        \includegraphics[width=\linewidth]{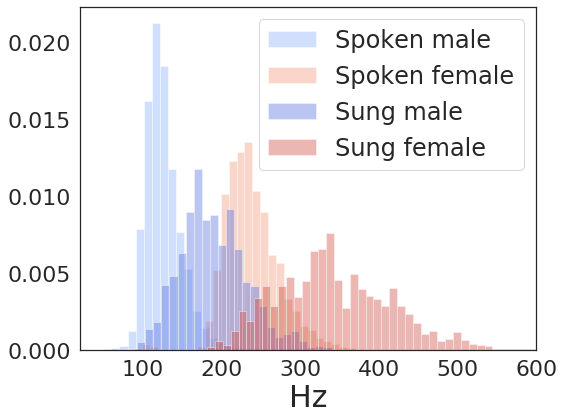}
        \caption{Vowels pitch values.}
        \label{fig:distPitch}
    \end{subfigure}
    \caption{Distribution of vowel duration (\ref{fig:distDuration}) and pitch value (\ref{fig:distPitch}), separated by sung and spoken speech styles.}
    \label{fig:Sung_vs_Read}
\end{figure}

First, we considered vowel duration differences (Figure \ref{fig:distDuration}). It is seen that whereas spoken vowel durations are typically less than 300 milliseconds (ms), sung vowels are often over 500 ms and can extend over 1 second.  Word recognition accuracy may be affected by the phone lengthening.  In \cite{Kawai2017}, it was found that the extended duration can lead to phone insertion and substitution errors, especially if the pitch is varying during the vowel. Further, vowel length is an important phonetic cue, (e.g., aiding discrimination between short and long vowel pairs such as /I/ vs /i:/). Artificially extending a vowel may therefore disrupt identification.

Figure \ref{fig:distPitch} shows the vowel's fundamental frequency (F0) distribution for speech style and gender.
Sung speech has a much larger F0 range than spoken speech.
In contrast to spoken speech, which has virtually no overlap in F0 range between genders, the extended F0 ranges mean that the top end of the male range (200-300 Hz) overlaps with the lower female F0 values.
For sung speech, different peaks can be seen in the male and female distributions.
These peaks occur at the position of specific musical notes, e.g., some of the peaks in the male singer plot correspond to the notes $E_3$, $F^{\#}_{3}$ and $G^{\#}_{3}$.
Notice that unlike spoken speech, which can be well represented with a single Gaussian in log frequency scale, the sung speech distribution would be better represented by a Gaussian mixture model where each component centred at a discrete note in the vocal range. At the highest sung F0 values vowel identification will become challenging as the sparseness of the harmonics means that the formant positions and bandwidths will be hard to determine.

%

\begin{figure}[!t]
    \centering
    \begin{subfigure}{.5\linewidth}
    \includegraphics[width=\linewidth]{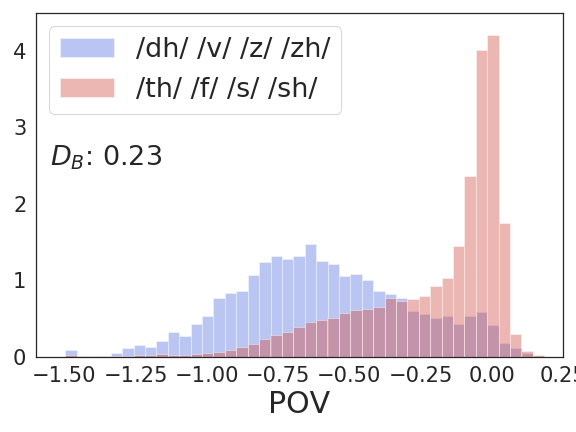}
        \caption{Sung speech.}
        \label{fig:povDistFricative_sung}
    \end{subfigure}%
    \begin{subfigure}{.5\linewidth}
    \includegraphics[width=\linewidth]{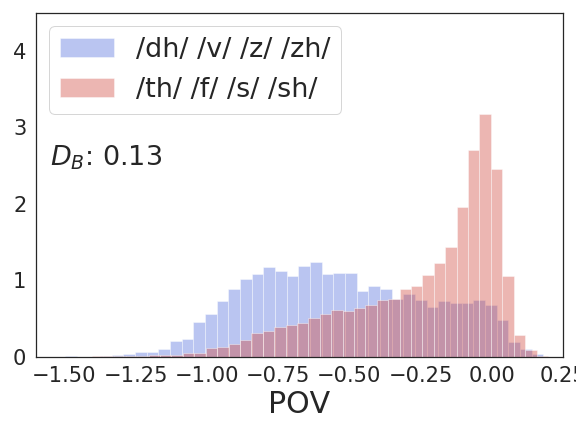}
        \caption{Spoken speech.}
        \label{fig:povDistFricative_read}
    \end{subfigure}
    \caption{Distribution of POV measure for voiced and unvoiced fricative phones. 
    Bhattacharyya distance ($D_B$) is annotated in both plots.}
    \label{fig:povDistFricative}
\end{figure}

We used the Kaldi `probability of voicing' measure (POV)  (Section \ref{sec:musicalfeatures}), to examine the importance of the voicing information for the discrimination between voiced and unvoiced phones. 
Figure \ref{fig:povDistFricative} shows the POV measure's probability distribution for the voiced/unvoiced fricative sounds (note, low POV implies more strongly voiced). 
For each condition, the Bhattacharyya distance ($D_B$) was measured to show the similarity between the voiced and unvoiced distributions. For sung speech, although both classes show more voicing (i.e., a shift of the distributions to the left), the shift is greater for voiced phonemes leading to a higher $D_B$ value. This suggests that the POV could be more informative for sung phonemes differentiation than for spoken speech. No variation on the $D_B$ distance was observed for voiced versus unvoiced stop phonemes. 

Finally, we analysed the voice quality (VQ) production differences between speech styles by using the fundamental frequency perturbance measurements jitter and shimmer,  plus harmonic to noise ratio (HNR); parameters typically used for voice pathology detection \cite{Teixeira2013}.
These features have been proved useful for speaker recognition \cite{Farrus2007}. 
VQ measurements are divided into three subgroups; jitter, that measures the frequency variation from cycle-to-cycle; shimmer, that relates to the amplitude variation of the sound wave; and HNR, that measures the ratio between the harmonics and the glottal noise.
Analysis using the NUS-48E showed that sung speech jitter and shimmer ranges are lower, and HNR is higher than in spoken speech.  
Among the factors that may explain these variations are vocal tract muscles "warm-up" during singing \cite{Mezzedimi2018}, interactions between jitter, shimmer and the degree of vibrato \cite{Hakes1988} and, the correlation between these parameters and the fundamental frequency, i.e., jitter will decrease at higher frequencies.
The voice quality parameters ranges showed a clear gender distinction in spoken speech.
In contrast, in singing, the gender distinction virtually disappears, only jitter maintains some degree of gender specification and, shimmer and HNR become more gender independent parameters. The weaker link to speaker identity may reduce the value of these parameters as speaker normalisation conditioning variables.

%% file: 3-pitchtracker.tex
\section{Voice Source Features}
\label{sec:musicalfeatures}

This section describes the voice source-based features that will be evaluated in the ASR experiments.

\subsection{Pitch and voicing estimation}

For pitch feature extraction, we utilised the Kaldi pitch tracker \cite{Ghahremani2014}, which is based on the getf0 (RAPT) algorithm \cite{talkin1995robust}. 
Unlike getf0, Kaldi pitch tracker does not make a hard voicing decision; instead, it treats all frames as voiced sounds with an associated pitch, but with varying probability of voicing (POV).
A Viterbi search is used to interpolate pitch estimates across unvoiced frames. 
The algorithm outputs a log pitch representation, and a normalised log pitch by using a short-time mean subtraction (as in \cite{lei2006modeling}), a Gaussian-distributed POV feature and a two frame-context delta-pitch feature.

\subsection{Tuning the Kaldi pitch tracker for sung speech}
\label{ssec:kaldiPitch}

The Kaldi pitch tracker possesses a maximum pitch value (max-f0) and low-pass frequency cut-off (lowpass-cutoff) that are tuned for spoken speech. These were re-tuned for sung speech using the MIR-1K pitch annotated sung speech dataset (MIR-1K) \cite{Hsu2010}.
MIR-1K is a collection of 1000 male and female song clips totalling 133 minutes, extracted from 110 karaoke songs selected from 5,000 Chinese pop songs. Pitch values have been manually annotated.

By employing a grid-searching technique, we evaluated seven values for max-f0 (between 400 and 1,000, with 100Hz step) and three values for lowpass-cutoff (1,000, 1,500 and 2,000). 
The performance was assessed by using the gross pitch error (GPE) \cite{Drugman2011} and fine pitch error (FPE) \cite{Asgari2013} measurements.
GPE calculates the proportion of frames classified as voiced by both the estimation and ground truth, where the estimated pitch deviates more than one semitone from the ground truth \cite{Babacan2013}.
However, as the Kaldi pitch tracker does not perform a hard voiced decision, we calculate the GPE using all the \textit{voiced frames} from the ground truth.
And, FPE is defined as the mean absolute error derived from the voiced frames where the reference deviates less than the GPE threshold \cite{Asgari2013}.

\begin{figure}
    \centering
    \includegraphics[width=\linewidth]{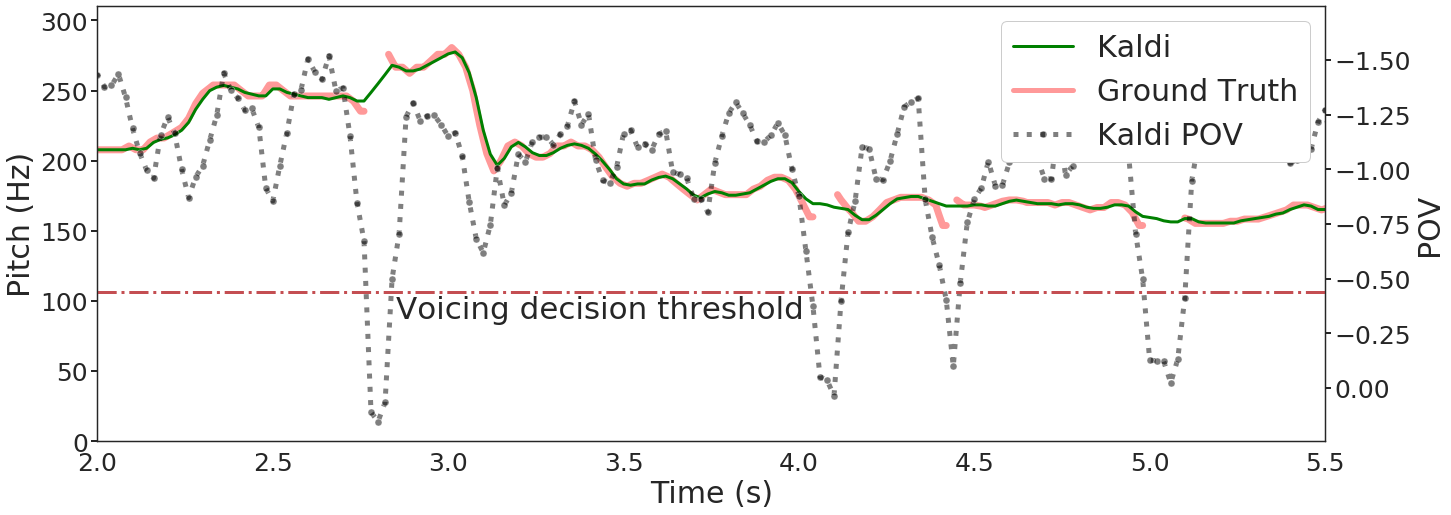}
    \caption{Kaldi pitch tracker's pitch and POV estimation contrasted with the ground truth from 3.5 seconds from one MIR-1K excerpt.}
    \label{fig:pitchtrack}
\end{figure}

Best performance was found with a max-f0 of 1,000Hz and lowpass-cutoff of 1,500Hz (compared to 400 and 1000 respectively for the original spoken speech tuning).
Figure \ref{fig:pitchtrack} shows an example of the predicted pitch by the Kaldi pitch tracker using these parameters, compared to the ground truth. 
Notice that Kaldi pitch tracker predicted the values with high accuracy and it interpolates between voiced frames to assign a pitch to the unvoiced ones. 
Additionally, this plot shows the POV estimation (right y-axis with inverted order), notice that in unvoiced areas, the POV value shows valleys to values below a voicing decision threshold estimated from MIR-1K.

\subsection{Voice quality parameters}

The VQ parameters were extracted using Praat \cite{praat}.
First, we estimated jitter by analysing the timings of individual fundamental periods.
In our experiments, we evaluated two jitter parameters, namely average absolute jitter (\textit{jitta}) and relative average perturbation (\textit{rap}) \cite{Teixeira2013}.
The \textit{jitta} parameter measures the average \textit{absolute} variation of the fundamental period between consecutive cycles, and \textit{rap} measures the \textit{relative} difference between a period and the average of its two neighbours.
Second, we estimated shimmer, defined as the average absolute difference between the \textit{amplitude} of pairs of consecutive periods.
Finally, HNR is a single parameter calculated from the auto-correlation function of the voice signal and is a measure of the degree of aperiodicity \cite{Teixeira2013}.

\begin{table*}[t]
    \centering
    \caption{Evaluation set performances (95\% confidence interval WER) from experiments trained on DSing1, DSing3 and DSing30. Values in bold show WERs that are significantly better than the baseline.}
    \footnotesize{
    \begin{tabular}{l|cc|cc|cc}
    \toprule
    \multirow{2}{*}{\textbf{Experiment}} & \multicolumn{2}{c}{DSing1} & \multicolumn{2}{c}{DSing3} & \multicolumn{2}{c}{DSing30} \\ 
                        & 3-gram & 4-gram & 3-gram & 4-gram & 3-gram & 4-gram\\
    \midrule
    Baseline &  
        $43.02\pm0.55$  & $38.14\pm0.58$ & $28.13\pm0.14$ & $24.40\pm0.26$ & $22.82\pm0.21$ & $19.88\pm0.34$ \\
    \midrule
    Kaldi LN  & 
        $\mathbf{40.99\pm0.49}$ & $\mathbf{36.77\pm0.45}$ & $28.05\pm0.24$          & $24.27\pm0.21$          & $23.23\pm0.28$ & $19.87\pm0.12$ \\
     
    Kaldi L  &  
        $\mathbf{41.55\pm0.58}$ & $37.14\pm0.52$          & $27.80\pm0.34$          & $24.35\pm0.27$          & $22.92\pm0.33$ & $19.72\pm0.25$ \\

    Kaldi N  &  
        $\mathbf{41.61\pm0.37}$ & $37.40\pm0.87$          & $\mathbf{27.79\pm0.27}$ & $24.28\pm0.19$          & $22.95\pm0.15$ & $19.67\pm0.11$ \\

    \midrule
    Kaldi LN + VQ &  
        $\mathbf{41.17\pm0.30}$ & $\mathbf{36.70\pm0.46}$ & $27.82\pm0.26$          & $\mathbf{23.76\pm0.27}$ & $22.97\pm0.32$ & $19.60\pm0.21$ \\

    \bottomrule
    \end{tabular}
    }
    \label{tab:ci}
\end{table*}

%% file: 4-baseline_system.tex
\section{The ASR baseline}
\label{sec:baseline}

\subsection{The task}

Recognition experiments have been performed using the DSing sung speech recognition corpus \cite{Roa2019}. DSing is composed of the 4,460 karaoke performances, and is the English subset of the larger multi-language karaoke Smule Sing!300x30x2 dataset (Sing!) \cite{DAMP} released by Smule\footnote{https://www.smule.com/}, in early 2018. 

The performances are equally distributed by gender and organised by the country (30 in total) where the singer was located during the recording.
Using the country information, DSing conveniently splits the data into three progressively larger training sets, named \textit{DSing1} (15.1 hours), \textit{DSing3} (44.7 hours) and \textit{DSing30} (149.1 hours). 
The smallest training set DSing1 is constructed using 80\% of the recordings from the UK.
The remaining 20\% is split into two balanced test sets, one for development (dev) and one for evaluation (eval) (1.5 hours in total).
DSing3 extends DSing1 with recordings from Australia and the US.
Although there is no information about the nationality of the singers, it can be safely assumed that most of the singers in DSing1 and DSing3 are native English speakers.
The third and most extensive set, DSing30, covers all the English recordings from the 30 countries, including a large number of English recordings from non-native English speakers.

\subsection{Baseline system description}
\label{ssec:baselinesystem}

As a baseline, we employ the sung speech recognition system from \cite{Roa2019} built using the Kaldi ASR toolkit \cite{Povey_ASRU2011}. The acoustic features employed are 13 Mel frequency cepstral coefficients (MFCC) plus delta, delta-delta and energy, with 25 ms frame length and 15 ms of overlapping. Initial alignments are performed with a tri-phone speaker adapted Gaussian mixture model (GMM) and feature-space maximum likelihood linear regression (fMLLR). The tri-phone model is used to clean the training data using the standard Kaldi cleanup process\footnote{https://github.com/kaldi-asr/kaldi/blob/master/egs/wsj/s5/steps/\\cleanup/clean\_and\_segment\_data.sh}. A factorised time-delay neural network (TDNN-F) \cite{Povey2018} acoustic model is then trained using 40 MFCCs with two frames context plus 100-dimensional i-Vectors, and a lattice-free maximum mutual information (LF-MMI) loss function \cite{Povey2016}.

A 3-gram MaxEnt language model (LM) is trained on an in-domain lyrics corpus, sourced from the Lyrics Wiki website\footnote{https://lyrics.fandom.com/wiki/LyricWiki}. This corpus is composed of the lyrics from songs from all artists in the DSing3 training set excluding songs that actually appear in DSing itself, plus the lyrics from all the artists from the Billboard The Hot 100 for the 31st December of the years 2015 to 2018.  Language model rescoring is then performed with a 4-gram MaxEnt model trained on the same data. The vocabulary selected was the 26K most frequency words.

Using the 3-gram LM, the evaluation set performance was 42.28\%, 28.67\% and 22.32\% for models trained with DSing1, DSing3 and DSing30 respectively, reducing to 37.63\%, 24.27\% and 19.60\% WER after applying 4-gram LM rescoring. For full system details see \cite{Roa2019}.

%% file: 5-experiments.tex
\section{Experiment and results}
\label{sec:experiments}
 
\subsection{Experiments}
In order to establish the significance of our results we have repeated experiments multiple times. In particular, the AM training depends on the \textit{random} initialisation of parameters, the \textit{random} presentation order of training data, etc. Each training, though equally valid, can produce WER results that vary appreciably, and the variation can be mistaken for genuine performance variations. Therefore, for all experiments we retrain systems 11 times allowing us to treat the evaluation of the model performances statistically.

We first replicated the baseline system training it eleven times to calculate confidence intervals by using the mean and the standard error of the mean scores. 
After each training, the development set was used to select the LM-weight and words-insertion-penalty Kaldi's decoding parameters, and these parameters were then used to decode the evaluation set.
These parameters were firstly estimated when decoding with the 3-gram LM and then when using the 4-gram LM for re-scoring.

Following the above procedure, we performed experiments by expanding the baseline MFCC + i-Vector feature vector using different voice source based feature combinations.
The first experiment, \textbf{Kaldi LN}, evaluated the effect of including both Kaldi pitch representations; the log pitch and the normalised log pitch.
The second, \textbf{Kaldi L}, evaluated the effect of using the log pitch, without the normalised log pitch.
The third, \textbf{Kaldi N}, evaluated the effect of utilising only the normalised log pitch.
For these three first experiments, both the delta pitch and POV were included.
The final experiment expanded the best combination of MFCC plus pitch features -- obtained from the results when training with the smallest DSing1 -- with the four \textbf{VQ} features.

\subsection{Results and analysis}

Table \ref{tab:ci} presents the system performances (WER) along with the 95\% confidence intervals for the baseline system and the systems using the various combinations of voice source features. For each system, results are shown for each training dataset and for using each of the 3-gram or 4-gram LM.
The baseline results for the 4-gram LM, $38.14\pm0.58$, $24.40\pm0.26$ and $19.88\pm0.34$ WER, for DSing1, DSing3 and DSing30 respectively, are consistent with the results reported in \cite{Roa2019}.

For the DSing1 model, using the four Kaldi pitch features (Kaldi LN experiment) reduces the error by about 2.0\% ($p<.05$) 3-gram and, 1.4\% ($p<.05$) 4-gram. A similar improvement is obtained by expanding Kaldi LN with the VQ features ($p<.05$).
For the larger DSing3 training set, a reduction of 0.5\% ($p=.057$) for 3-gram and, 0.7\% ($p<.05$) 4-gram, is obtained when combining the pitch feature with the VQ.
Voiced source features did not help to improve the models trained on the largest DSing30. The VQ features do not produce a significant improvement over using the pitch features, for any training set size.



\begin{figure}[!t]
    \centering
    \includegraphics[width=\linewidth]{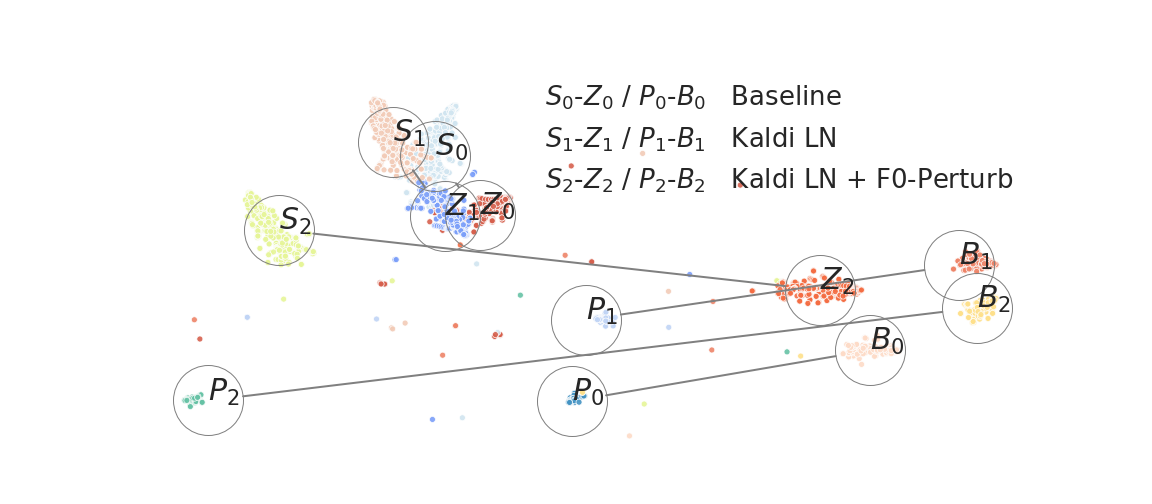}
    \caption{T-SNE constructed with the posterior probabilities from model train on DSing30.}
    \label{fig:tsne}
\end{figure}

Figure \ref{fig:tsne} shows a t-SNE plot of the \textbackslash{s}\textbackslash-\textbackslash{z}\textbackslash fricative and \textbackslash{p}\textbackslash-\textbackslash{b}\textbackslash plosive sounds, constructed by using the posterior probabilities from one model trained with DSing30.
The subscript 0 represents the baseline model; subscript 1 Kaldi LN and, subscript 2 Kaldi LN + VQ.
In the baseline model there are separate clusters for the voiced and unvoiced fricatives and plosive pairs, but the clusters overlap.   
When the model is informed with pitch features and VQ information, the distance between classes becomes wider and the classes are more compressed.
Best discrimination is seen when the VQ features are included, giving some hint that they may be useful even though improvements in WER were not significant. 

%% file: 6-conclusion.tex
\section{Conclusions}
\label{sec:conclusion}

This paper presented an analysis of the differences in voice source characteristic between the sung and spoken speech styles, and its effect on a sung speech ASR performance. Using voice source features improved ASR performances, reducing WERs by between 0.7\% to 1.4\% on an existing state of the art baseline.  
The ASR results and t-SNE analysis suggested that voice source features help to compress the phoneme classes and to increase the distance between pairs of voiced/unvoiced phonemes.
However, this effect only improves performances for ASR systems with the smaller training sets. When using the more extensive DSing30 (149.1 hrs) the vocal source based features were, surprisingly, found to provide no significant benefit for ASR performance. This suggests that, with enough training data sets, systems are able to learn the phonetic cues being carried in the voice source in a less direct manner (e.g., via the temporal dynamics of the MFCC features).

